\documentclass[11pt]{article}
\usepackage{graphicx}
\usepackage{amssymb,amsmath,amsfonts}

\def\Re{{\cal R \mskip-4mu \lower.1ex \hbox{\it e}\,}}
\def\Im{{\cal I \mskip-5mu \lower.1ex \hbox{\it m}\,}}
\def\ie{{\it i.e.}}

\def\sub#1{_{\lower.25ex\hbox{$\scriptstyle#1$}}}
\def\tev{\,{\ifmmode\mathrm {TeV}\else TeV\fi}}
\def\gev{\,{\ifmmode\mathrm {GeV}\else GeV\fi}}
\def\mev{\,{\ifmmode\mathrm {MeV}\else MeV\fi}}
\def\mpl{\ifmmode M_{pl}\else $M_{pl}$\fi}
\def\mpl{\ifmmode \overline M_{Pl}\else $\bar M_{Pl}$\fi}
\def\to{\rightarrow}

\def\subw{_{\rm w}}
\def\mh{\ifmmode m\sbl H \else $m\sbl H$\fi}
\def\mch{\ifmmode m_{H^\pm} \else $m_{H^\pm}$\fi}
\def\mt{\ifmmode m_t\else $m_t$\fi}
\def\mc{\ifmmode m_c\else $m_c$\fi}
\def\mz{\ifmmode M_Z\else $M_Z$\fi}
\def\mw{\ifmmode M_W\else $M_W$\fi}
\def\mws{\ifmmode M_W^2 \else $M_W^2$\fi}
\def\mhs{\ifmmode m_H^2 \else $m_H^2$\fi}   
\def\mzs{\ifmmode M_Z^2 \else $M_Z^2$\fi}
\def\mts{\ifmmode m_t^2 \else $m_t^2$\fi}
\def\mcs{\ifmmode m_c^2 \else $m_c^2$\fi}
\def\mchs{\ifmmode m_{H^\pm}^2 \else $m_{H^\pm}^2$\fi}
\def\ztwo{\ifmmode Z_2\else $Z_2$\fi}
\def\zone{\ifmmode Z_1\else $Z_1$\fi}
\def\mtwo{\ifmmode M_2\else $M_2$\fi}
\def\mone{\ifmmode M_1\else $M_1$\fi}
\def\tb{\ifmmode \tan\beta \else $\tan\beta$\fi}
\def\xw{\ifmmode x\subw\else $x\subw$\fi}
\def\ch{\ifmmode H^\pm \else $H^\pm$\fi}
\def\lum{\ifmmode {\cal L}\else ${\cal L}$\fi}
\def\inpb{\,{\ifmmode {\mathrm {pb}}^{-1}\else ${\mathrm {pb}}^{-1}$\fi}}
\def\infb{\,{\ifmmode {\mathrm {fb}}^{-1}\else ${\mathrm {fb}}^{-1}$\fi}}
\def\epem{\ifmmode e^+e^-\else $e^+e^-$\fi}
\def\ppb{\ifmmode \bar pp\else $\bar pp$\fi}
\def\bsg{\ifmmode B\to X_s\gamma\else $B\to X_s\gamma$\fi}
\def\bsll{\ifmmode B\to X_s\ell^+\ell^-\else $B\to X_s\ell^+\ell^-$\fi}
\def\bstt{\ifmmode B\to X_s\tau^+\tau^-\else $B\to X_s\tau^+\tau^-$\fi}
\def\lamt{\ifmmode \tilde\lambda\else $\tilde\lambda$\fi}
\def\shat{\ifmmode \hat s\else $\hat s$\fi}
\def\that{\ifmmode \hat t\else $\hat t$\fi}
\def\uhat{\ifmmode \hat u\else $\hat u$\fi}

\newskip\zatskip \zatskip=0pt plus0pt minus0pt
\def\matth{\mathsurround=0pt}
\def\lsim{\mathrel{\mathpalette\atversim<}}
\def\gsim{\mathrel{\mathpalette\atversim>}}
\def\atversim#1#2{\lower0.7ex\vbox{\baselineskip\zatskip\lineskip\zatskip
  \lineskiplimit 0pt\ialign{$\matth#1\hfil##\hfil$\crcr#2\crcr\sim\crcr}}}

\def\grtsim{\,\,\rlap{\raise 3pt\hbox{$>$}}{\lower 3pt\hbox{$\sim$}}\,\,}
\def\lsim{\,\,\rlap{\raise 3pt\hbox{$<$}}{\lower 3pt\hbox{$\sim$}}\,\,}

\renewcommand{\thefootnote}{\fnsymbol{footnote}}

\begin{document} \begin{titlepage}
\rightline{\vbox{\halign{&#\hfil\cr
%&DRAFT\cr
&SLAC-PUB-13187\cr
%&March 2006\cr
}}}
\begin{center}
\thispagestyle{empty} \flushbottom { {
\Large\bf Z' Coupling Information from the LHeC  
\footnote{Work supported in part
by the Department of Energy, Contract DE-AC02-76SF00515}
\footnote{e-mail:
rizzo@slac.stanford.edu}}}
\medskip
\end{center}

\centerline{Thomas G. Rizzo}
\vspace{8pt} 
\centerline{\it Stanford Linear
Accelerator Center, 2575 Sand Hill Rd., Menlo Park, CA, 94025}

\vspace*{0.3cm}

\begin{abstract}
\end{abstract}
If the LHC discovers a $Z'$-like state the extraction of its couplings to the particles of the Standard Model becomes 
mandatory in order to determine the nature of the underlying new physics theory. It has been well-known for some time that the direct 
measurements performed at the LHC in the Drell-Yan channel cannot determine these parameters uniquely in a model-independent manner 
even if large integrated luminosities, $\sim 100 fb^{-1}$, 
become available and the $Z'$ is relatively light $\lsim 1.5$ TeV. Here we examine the possibility that a proposed $e_{L,R}^\pm p$ 
collider upgrade at the LHC, the LHeC, with $\sqrt s=1.5-2$ TeV could be helpful with such coupling determinations in the years 
before a Linear Collider is constructed. We show that the polarization and charge asymmetries constructed from the cross 
sections for these processes can be useful in this regard depending upon the specific values of the particular $Z'$ model parameters. 
%\vskip0.45in
%\begin{center}

%\end{center}

\renewcommand{\thefootnote}{\arabic{footnote}} \end{titlepage} 

%
%
%
%%%%%%%%%%%%%%%%%%%%%%%%%%%%%%%---- Put text here

\section{Introduction and Background}

The LHC turns on later this year and will begin a detailed study of the Terascale. Amongst the many prospective  
new physics signals that may be realized at the LHC the possibility of a $Z'$-like object appearing in the Drell-Yan channel is a 
relatively common one{\cite {reviews}} as it can arise in many scenarios. Due to the cleanliness of this channel, if such a state 
is light enough, it might be 
the first new physics to be discovered at the LHC even with relatively low integrated luminosities{\cite {ATLASTDR,CMSTDR}}. Once 
such a particle is found and its mass, width 
and spin are determined, we will want to know how it couples to the various fields of the Standard Model(SM) and in this 
way learn about the nature of the underlying new physics model. Over the years a number of authors have shown{\cite {reviews}} 
that observables are available at the LHC by which certain combinations of these couplings to the SM fermions can be determined in a model-independent 
manner.  
These combinations can be measured with reasonable precision by both ATLAS and CMS provided sufficient integrated luminosity, $\gsim 100 fb^{-1}$, 
is available and the new $Z'$ state is not too massive, $\lsim 1.5$ TeV. These results have recently been shown to remain valid 
even when a full Next-to-Leading Order(NLO) analysis is performed{\cite {frank}}. 

One of the main issues that these studies have uncovered is that there is insufficient information available 
to make completely model-independent extractions of {\it all} of the desired $Z'$ couplings. One answer to this dilemma is to wait until 
we have available a linear collider in the 500 GeV/1 TeV (or more) energy range. As is well-known{\cite {ilc}}, it is 
possible to extract $Z'$ coupling information at such colliders by indirectly observing the virtual effects 
of $Z'$ exchange below the direct production threshold. Of course a sufficiently energetic linear collider may even allow us 
to sit on the $Z'$ resonance peak, repeating the LEP/SLC experience, thus obtaining all of the interesting couplings with high precision. 
It seems likely, however, that such a machine may 
not be constructed until at least the mid-to-late 2020's, which is a long time to wait for such valuable additional information on new physics 
discoveries made at the LHC. 

There may, however, be another way to obtain indirect yet independent coupling information on a much faster time scale. 
There has been a proposal{\cite {lhec}} to generate $e_{L,R}^{\pm}p$ collisions using the 7 TeV proton beams from the LHC 
together with either an additional electron ring in the LHC tunnel or by employing a new $\sim 6$ km electron linac. 
If such a device, termed the LHeC, were 
to be constructed, $e^\pm p$ collisions would take place in the $\sqrt s=1.4\sim 2$ TeV energy range with an integrated 
luminosity of order $\sim 100 fb^{-1}$ and would allow for possible $e^\pm$ beam polarization. Such collisions might then be 
used to extract coupling information from the already discovered $Z'$ for a certain ranges of masses and model parameters. The purpose of this 
paper is to make a preliminary exploratory 
estimate of the potential sensitivity of such measurements performed at an LHeC for a $Z'$ in the mass range indicated above.  
Much improved estimates, of course, will become possible once the detailed parameters of the proposed LHeC become more definitive; the 
discovery of a light $Z'$ may then provide additional motivation to construct such a device. .

\section{Analysis}

The possibility of sensitivity to $Z'$ exchange in high-energy $e^\pm p$ collisions is not new{\cite {old}}
and dates back many years. Following a modified version of the notation employed by Capstick and Godfrey{\cite {steve}}, the differential 
cross section for polarized $e_L^\pm p$ deep inelastic scattering in the neutral current channel can be written as
\begin{eqnarray}
{{d\sigma(e^-_Lp)}\over {dxdy}} &=& {{2\pi \alpha^2}\over {sxy^2}}\sum_q 
\Big(q(x,Q^2)~[b_{LL}^2+b_{LR}^2(1-y)^2]+\bar q(x,Q^2)~[b_{LR}^2+b_{LL}^2(1-y)^2]
\Big)\\ \nonumber
{{d\sigma(e^+_Lp)}\over {dxdy}} &=& {{2\pi \alpha^2}\over {sxy^2}}\sum_q 
\Big(q(x,Q^2)~[b_{RL}^2+b_{RR}^2(1-y)^2]+\bar q(x,Q^2)~[b_{RR}^2+b_{RL}^2(1-y)^2]
\Big)\,,
\end{eqnarray}
where $q$ labels the various flavors of partons in the initial state with $q(\bar q)(x,Q^2)$ being the relevant 
parton densities(PDF) and $Q^2=sxy$, with $x,y$ being the traditional deep inelastic scattering variables. For the corresponding 
case of $e_R^\pm p$ scattering, we simply interchange all of the $R$ and $L$ labels everywhere in the expressions above. 
Furthermore, one finds that the $b_{ij}$, with $i,j$ being a helicity labels, are given by 
\begin{equation}
b_{ij}=\sum_n g_i^e(n)g_j^q(n)~{{Q^2}\over {Q^2+M_n^2}}\,,
\end{equation}
with $n$ labeling the masses and couplings of the the various gauge bosons exchanged in the $t-$channel, 
\ie, $n=\gamma, Z, Z'$, and $g_i^{e,q}(n)$ are the relevant chiral couplings of the electron and quarks to 
these gauge bosons in units of proton charge $e$. Note that since no resonances appear in this channel it is safe to 
neglect the finite widths of the $Z$ and $Z'$ in the present analysis. Obviously for the case of $M_{Z'}\geq 1$ TeV as considered here,
very large $Q^2$ will be necessary to show any statistically significant non-SM behavior. From these expressions it is clear that the 
shifts in the cross sections due to a $Z'$ of a known fixed mass will depend upon the eight {\it products} of unknown 
couplings, $g_{L,R}^e(Z')g_{L,R}^{u,d}(Z')$; here we have assumed that these couplings are generation-independent as in many, but not all, models 
that predict the existence of a $Z'$. It is just these specific  
coupling combinations that we would hope to be able to constrain by fits to this deep inelastic scattering data. Certainly, the individual 
$Z'$ couplings themselves cannot be determined from these measurements alone but one can then combine them with the more familiar ones 
obtained from the LHC in the Drell-Yan (and possibly other) channel(s). Note that if the gauge group, $G$, which contains the generator to which the 
$Z'$ under discussion couples, commutes with the $SU(2)_L$ of the SM then $g_L^{u}(Z')=g_L^{d}(Z')$ and we have only six combinations 
of independent couplings to determine. This is a rather common feature of many $Z'$ models{\cite {reviews}}.

One could now go ahead and calculate the four differential quantities $d\sigma(e^\pm_{L,R}p)$ and compare the predictions 
of various models with $Z'$ exchanges included to those of the SM and to each other. There are three reasons not to follow this direct approach. First, 
as is well-known, overall cross section measurements suffer from luminosity uncertainties. Though we might expect these to 
be rather small in this channel, $\sim O(5\%)$, we are at the same time looking to examine small deviations in these cross sections so it 
is best to avoid this problem if possible even though it is only an overall normalization uncertainty. Second, as we are reminded by Ref.{\cite {perez}}, 
the large $x$ 
parts of these fully differential cross sections which will show the greatest $Z'$ sensitivity are also those that have the greatest 
PDF uncertainties. Lastly, we might expect that NLO corrections to the individual cross sections to be comparable to the effects we are examining.     
To this end we note that one can define a number of asymmetries{\cite {old}}, which being the ratios of 
differential cross sections, suffer far less due to either of the above uncertainties as well as to the NLO corrections and we will employ these in our 
analysis below. (This argument does not, of course, imply that the cross sections themselves cannot be used for these purposes but only that 
additional care would be required in performing such an analysis.) Given the four polarized cross sections we can form the six polarization-dependent 
asymmetries:
\begin{eqnarray}
A^\pm &=& {{d\sigma(e_L^\pm)-d\sigma(e_R^\pm)}\over {d\sigma(e_L^\pm)+d\sigma(e_R^\pm)}}\\ \nonumber
C_{L,R} &=& {{d\sigma(e_{L,R}^-)-d\sigma(e_{L,R}^+)}\over {d\sigma(e_{L,R}^-)+d\sigma(e_{L,R}^+)}}\\ \nonumber 
B_{1,2} &=& {{d\sigma(e_{L,R}^-)-d\sigma(e_{R,L}^+)}\over {d\sigma(e_{L,R}^-)+d\sigma(e_{R,L}^+)}}\,.
\end{eqnarray}
The first pair of these is just the left-right asymmetries for each charge state while the second pair corresponds to 
the polarization-dependent charge 
asymmetries; the third pair is the mixed asymmetries. Of course, only four of these quantities are independent since there are only four 
differential cross sections to begin with but we will consider 
the full set here as each individual asymmetry displays somewhat different $Z'$ sensitivities as will be seen below. 
If polarization is not available for some reason then the ordinary charge asymmetry can always be employed: 
\begin{equation}
A_Q={{d\sigma(e^-)-d\sigma(e^+)}\over {d\sigma(e^-)+d\sigma(e^+)}}\,,
\end{equation}
where a sum over polarizations is implied. 
These asymmetries will clearly depend upon different combinations of the $Z'$ couplings than do the usual LHC $Z'$ observables thus 
providing additional model constraints that can be used in a simultaneous fit. It is important to remember that the mass of the 
$Z'$ will be well determined at the time that these asymmetries are measured at the LHeC and can be used as input; we will make 
this assumption in the numerical analyses performed below.

\begin{figure}[htbp]
\centerline{
\includegraphics[width=5.7cm,angle=90]{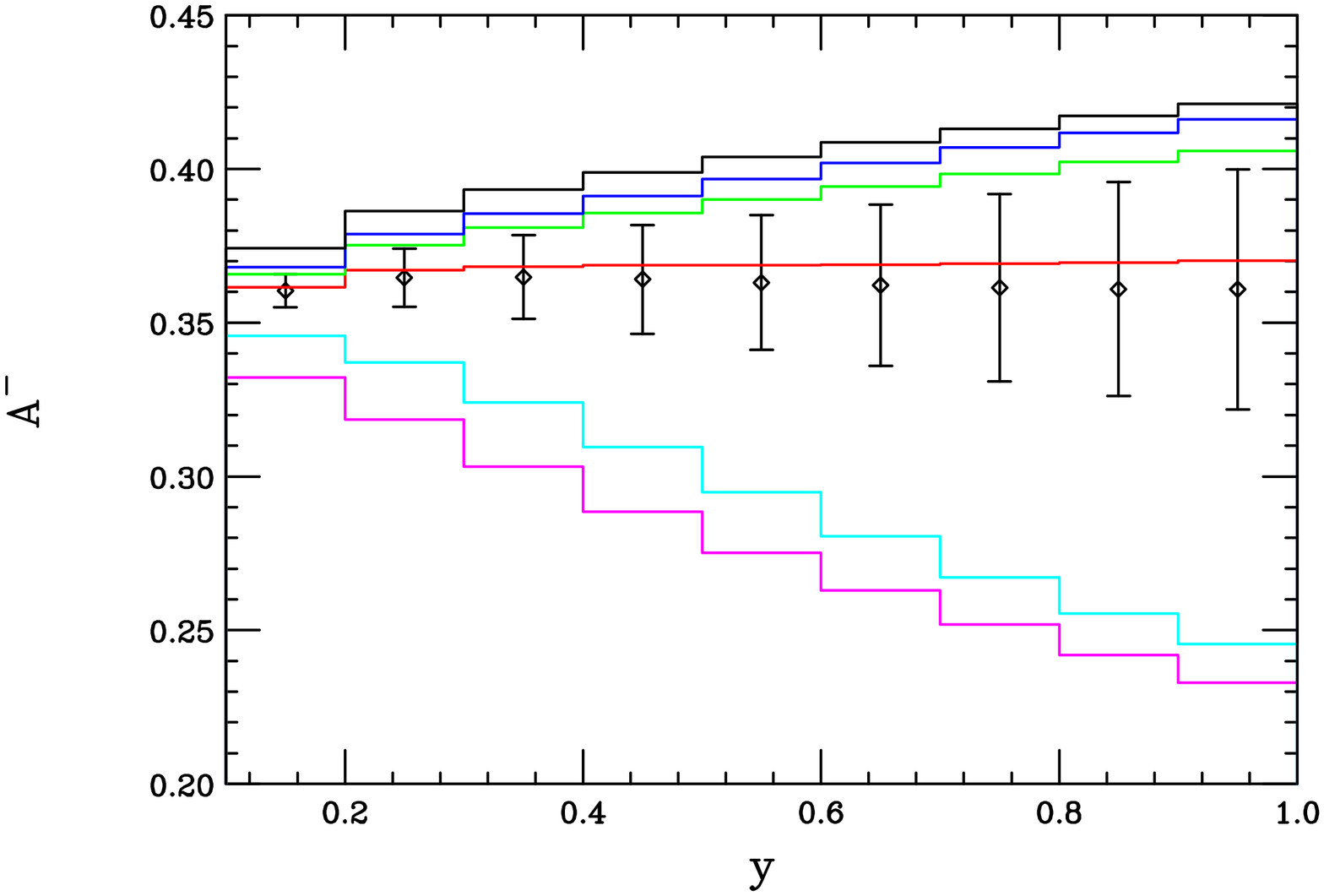}
\hspace*{0.1cm}
\includegraphics[width=5.7cm,angle=90]{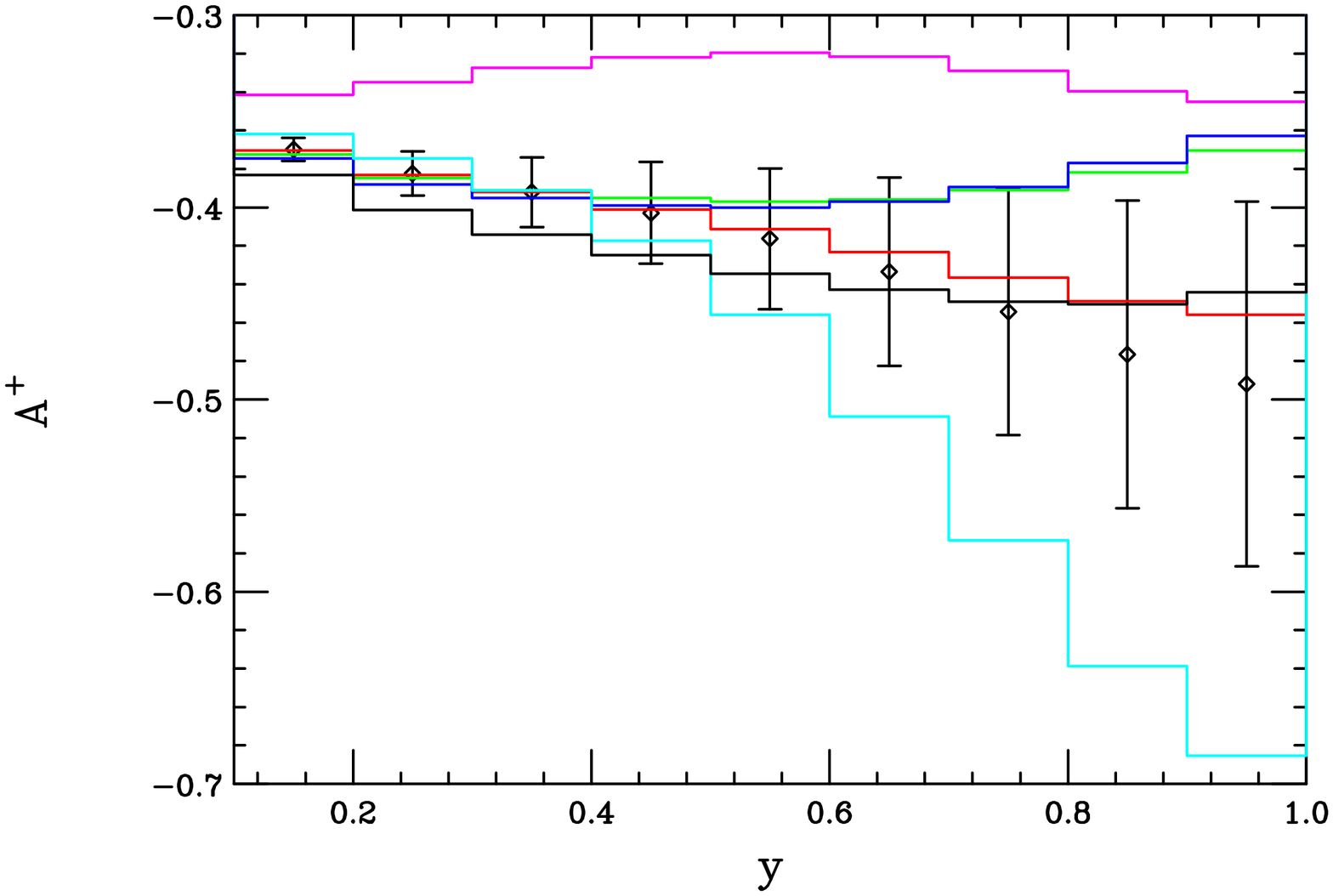}}
\vspace*{0.1cm}
\centerline{
\includegraphics[width=5.7cm,angle=90]{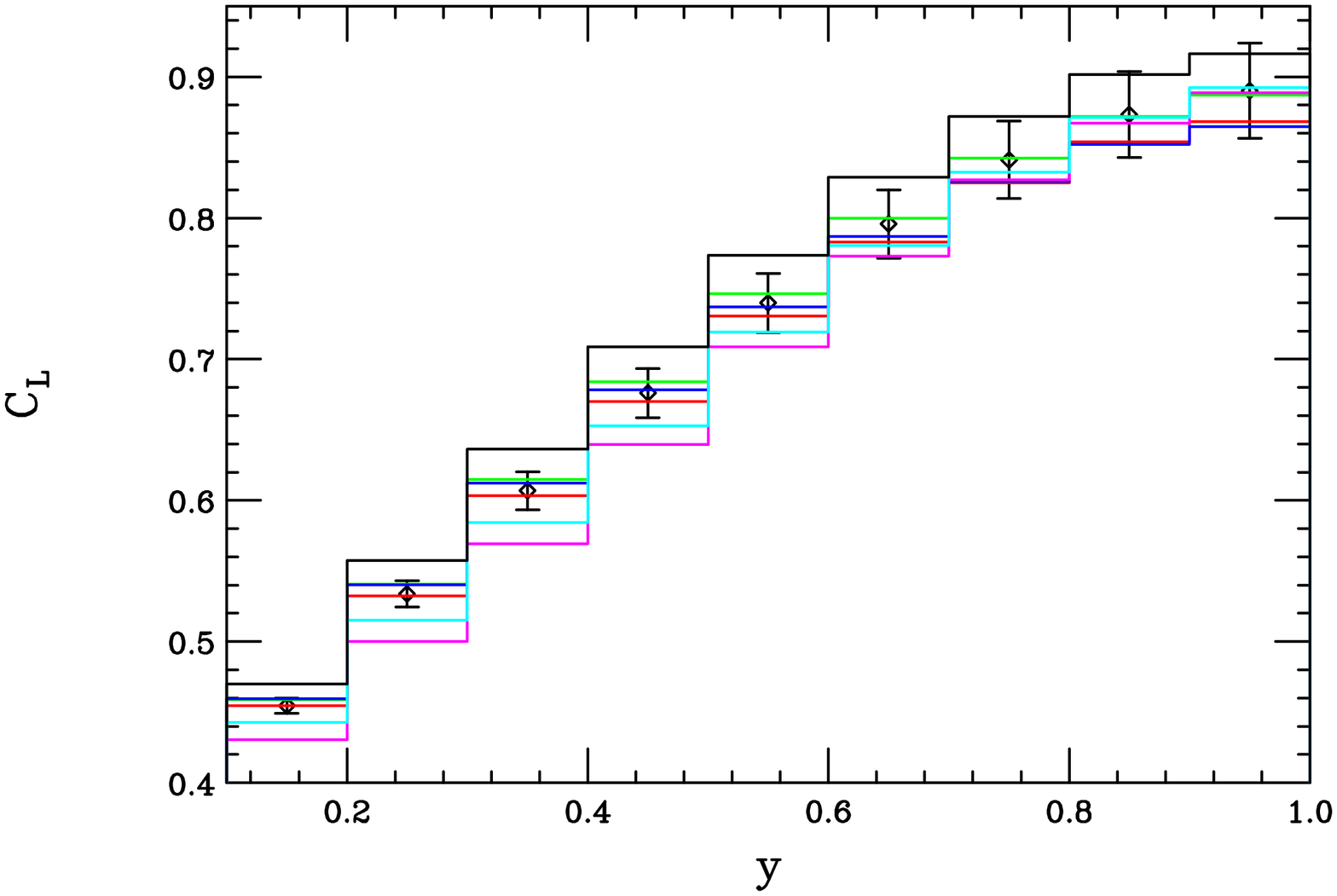}}
\caption{The asymmetries $A^\pm,C_L$ as functions of $y$ in 0.1 bins 
as described in the text assuming $\sqrt s=1.5$ TeV and $M_{Z'}=1.2$ 
TeV. The data points are the SM predictions with their associated errors. The red(green,blue,magenta,cyan,black) histograms 
correspond to the predictions of the $\psi(\chi,\eta$,LRM,ALRM,SSM) $Z'$ models, respectively.}
\label{fig1}
\end{figure}

When making comparisons with the predictions of the SM we will employ only statistical errors which likely dominate for these  
asymmetry measurements. We will make the further simplifying assumption that all of the $e_{L,R}^\pm$ channels have the 
same integrated luminosity and degree of polarization, $P$. Of course in a more realistic situation this will not likely be true as 
one might expect somewhat lower luminosity and polarization for initial state positrons. These effects can be evaluated 
once a more detailed 
accelerator design is on hand.  However, in that idealized limit, denoting the set of six asymmetries above generically 
by $A_{ij}$, the relevant error is essentially given by
\begin{equation}
\delta A_{ij}= \Big({{1-(PA_{ij})^2}\over {(N_i+N_j)P^2}}\Big)^{1/2}\,,
\end{equation}
where $N_{i,j}$ are the relevant number of events in the two channels. For the case of the charge asymmetry the associated 
error is given by a somewhat similar expression with $P=1$ and the total event rate then appears in the denominator. Once the LHeC 
proposal becomes more definitive these errors can be more realistically evaluated; our purpose here is only to give an indication 
of what may be possible at such a collider.  
For numerical purposes in the analysis we will assume a common value of $P=80\%$ as well as 
an integrated luminosity of $100 fb^{-1}$ in our discussion below. The results for other luminosities and polarizations can 
then be found in most cases by a simple scaling of the errors we obtain.

\begin{figure}[htbp]
\centerline{
\includegraphics[width=5.7cm,angle=90]{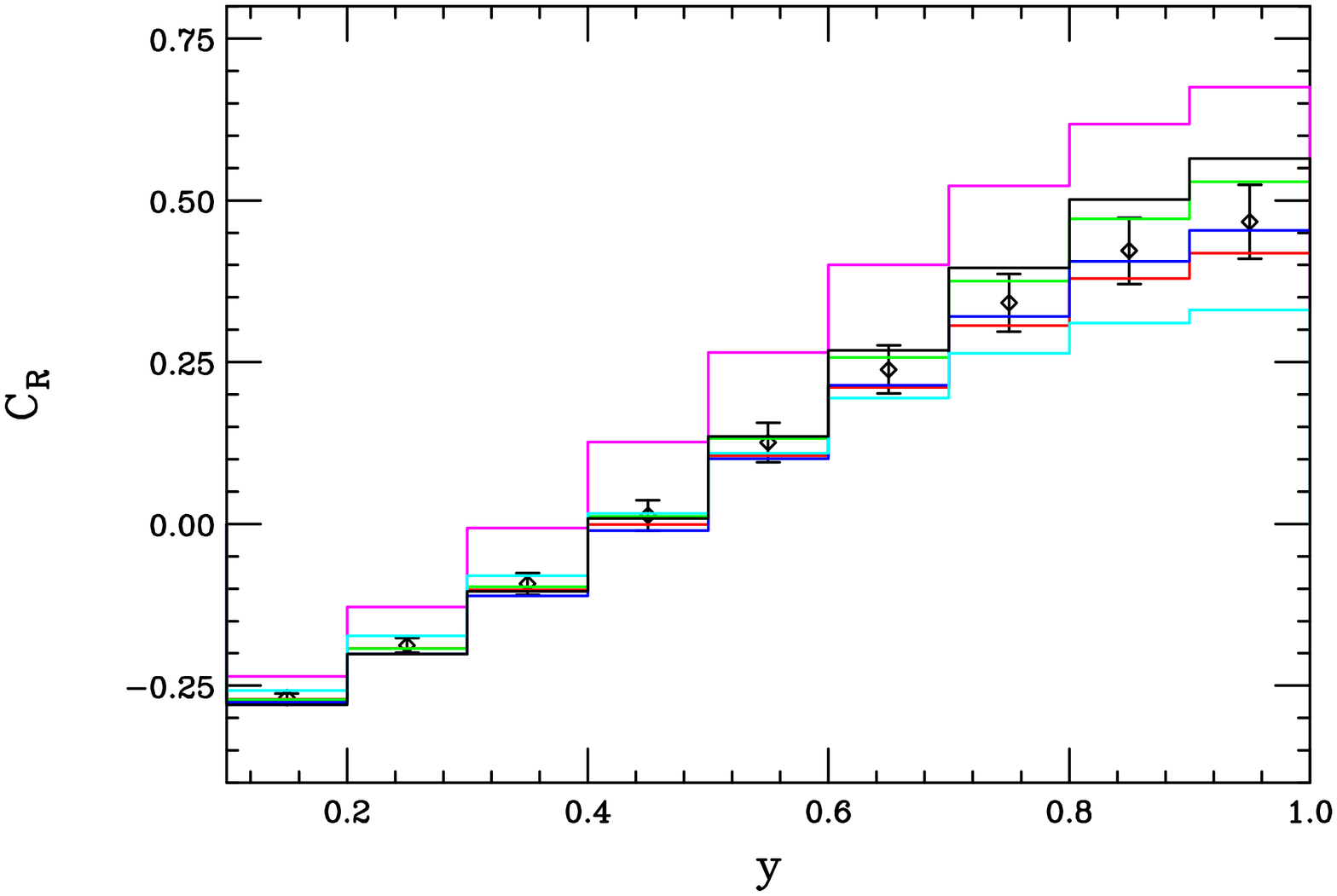}
\hspace*{0.1cm}
\includegraphics[width=5.7cm,angle=90]{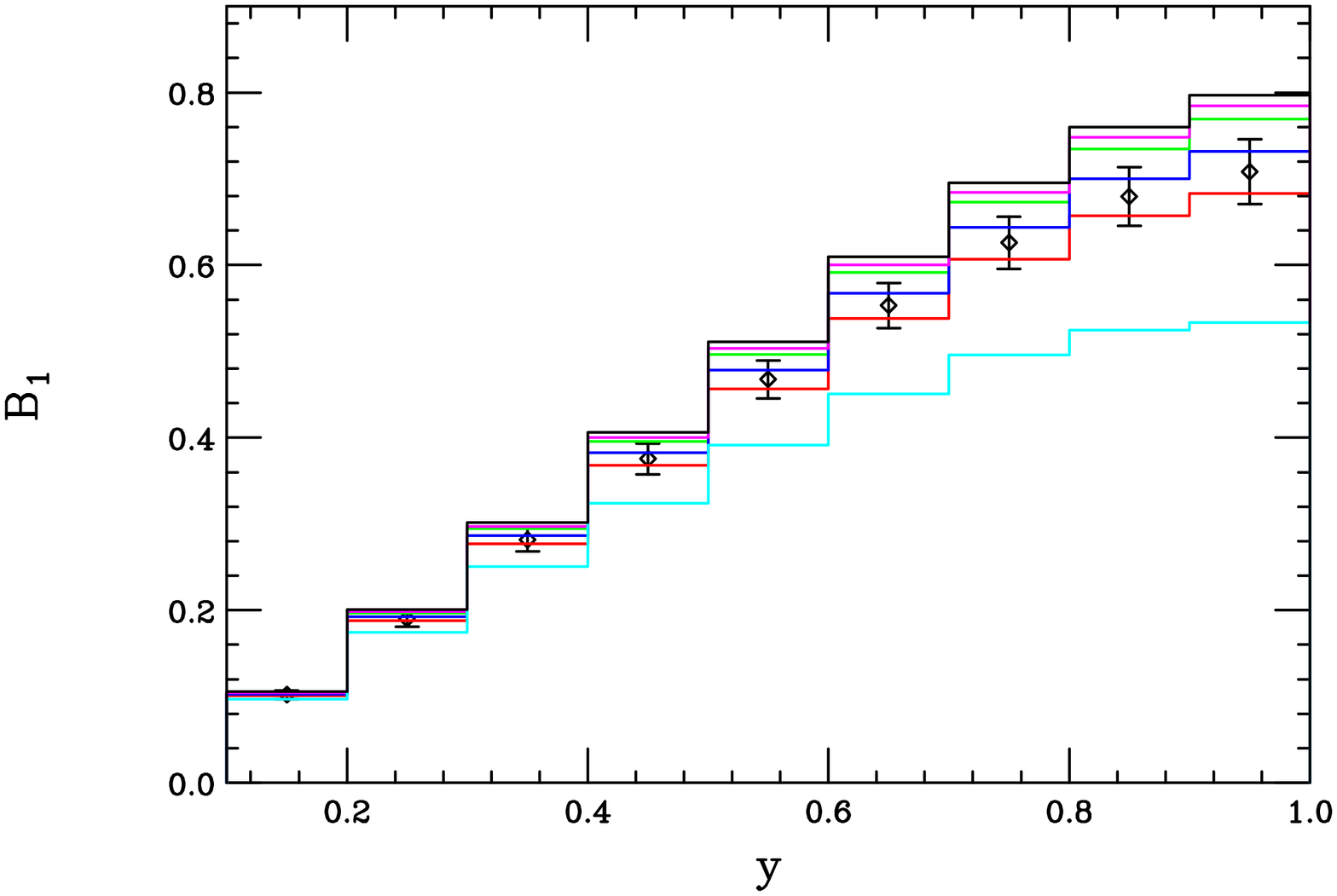}}
\vspace*{0.2cm}
\centerline{
\includegraphics[width=5.7cm,angle=90]{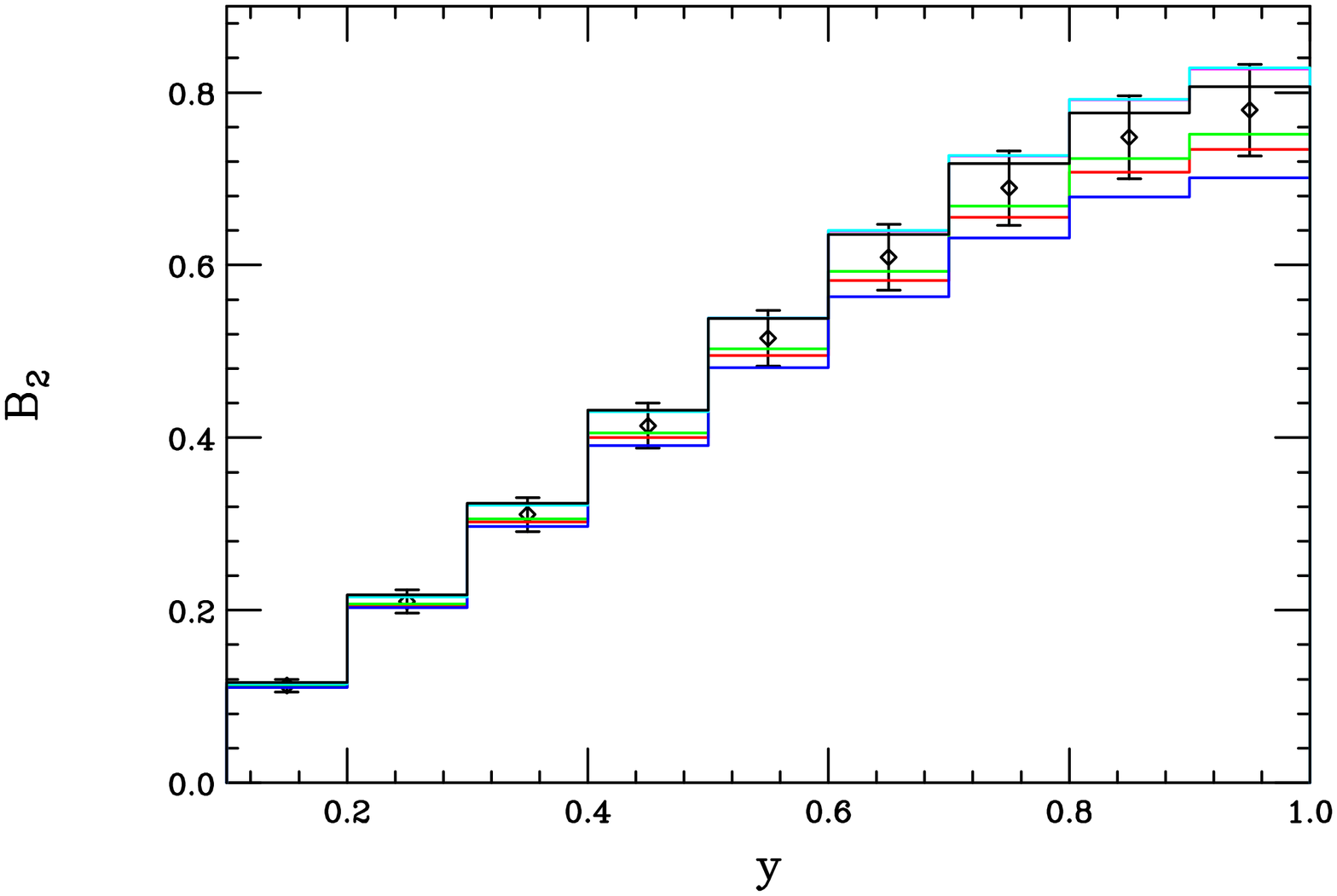}}
\caption{Same as the previous figure but now for the asymmetries $C_R,B_{1,2}$.}
\label{fig2}
\end{figure}
\begin{figure}[htbp]
\centerline{
\includegraphics[width=5.7cm,angle=90]{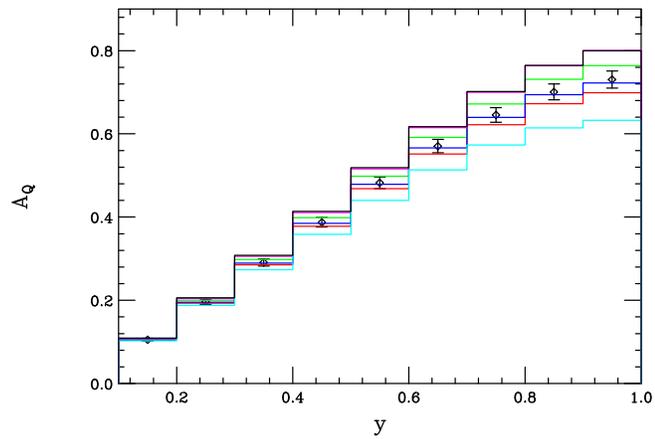}}
\caption{Same as the previous figure but now for the $A_Q$ asymmetry.}
\label{fig3}
\end{figure}

Since we wish to probe the largest $Q^2$ range for a given value of $\sqrt s$ and thus increase the $Z'$ sensitivity, 
we will require both $0.25 \leq x \leq 1$ (which we will then integrate over) and $0.1 \leq y \leq 1$; we then calculate the 
relevant asymmetries as a function of $y$ in the allowed range. These cuts, assuming $\sqrt s=1.5[2]$ TeV, removes the very small 
$Q^2>(237 GeV)^2[(316 GeV)^2]$ range which is totally dominated by the SM exchange contributions and which has been 
already been explored in detail at HERA{\cite {hera}}.

In order to gauge the sensitivity to new $Z'$ couplings we will consider and compare three possible scenarios for the combinations 
of the LHeC $\sqrt s$ and the mass $M_{Z'}$.  
As a first sample analysis, let us consider the case of $\sqrt s=1.5$ TeV assuming a $Z'$ mass of 1.2 TeV. In order to display the 
model dependency, we will consider the following small but generally familiar set of model scenarios{\cite {reviews}}: the GUT-based 
$E_6$ models $\psi,\chi$ 
and $\eta$, the Left-Right and Alternative Left-Right Models (LRM and ALRM, respectively) and the sequential Standard 
Model (SSM). Note that the present lower bounds on the masses of the $Z'${\cite {moriond}} in the above models at the Tevatron 
comes from CDF, based on $2.5 fb^{-1}$ of integrated luminosity, presently lie in the range of 853-966 GeV. The 
results for the various asymmetries in this sample case are shown in Figs.~\ref{fig1},~\ref{fig2} and ~\ref{fig3}. The first things 
we notice 
are that all of the asymmetries display reasonable model sensitivity and that the predictions of model $\psi$ in all cases 
cannot be distinguished from those of the SM due to its rather reduced couplings. The different asymmetries clearly show varying sensitivities 
to the choice of $Z'$ model; many of the models give systematically high or low asymmetry values for all $y$ in comparison to the predictions 
of the SM. 
Furthermore, if our error estimates are even approximately correct it is clear that significant new information about the $Z'$ 
couplings will be obtainable from the data under the above assumptions.

\begin{figure}[htbp]
\centerline{
\includegraphics[width=5.7cm,angle=90]{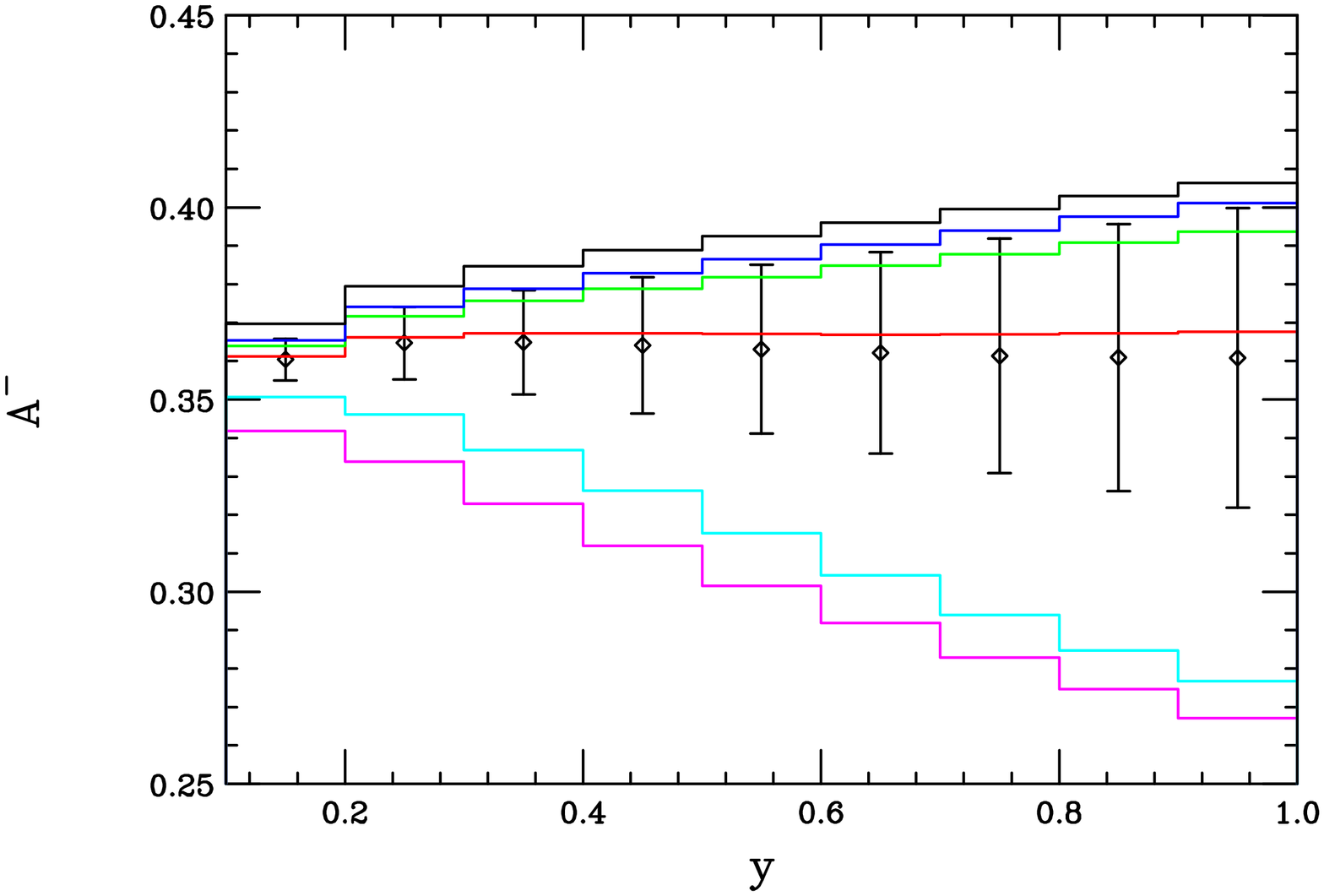}
\hspace*{0.1cm}
\includegraphics[width=5.7cm,angle=90]{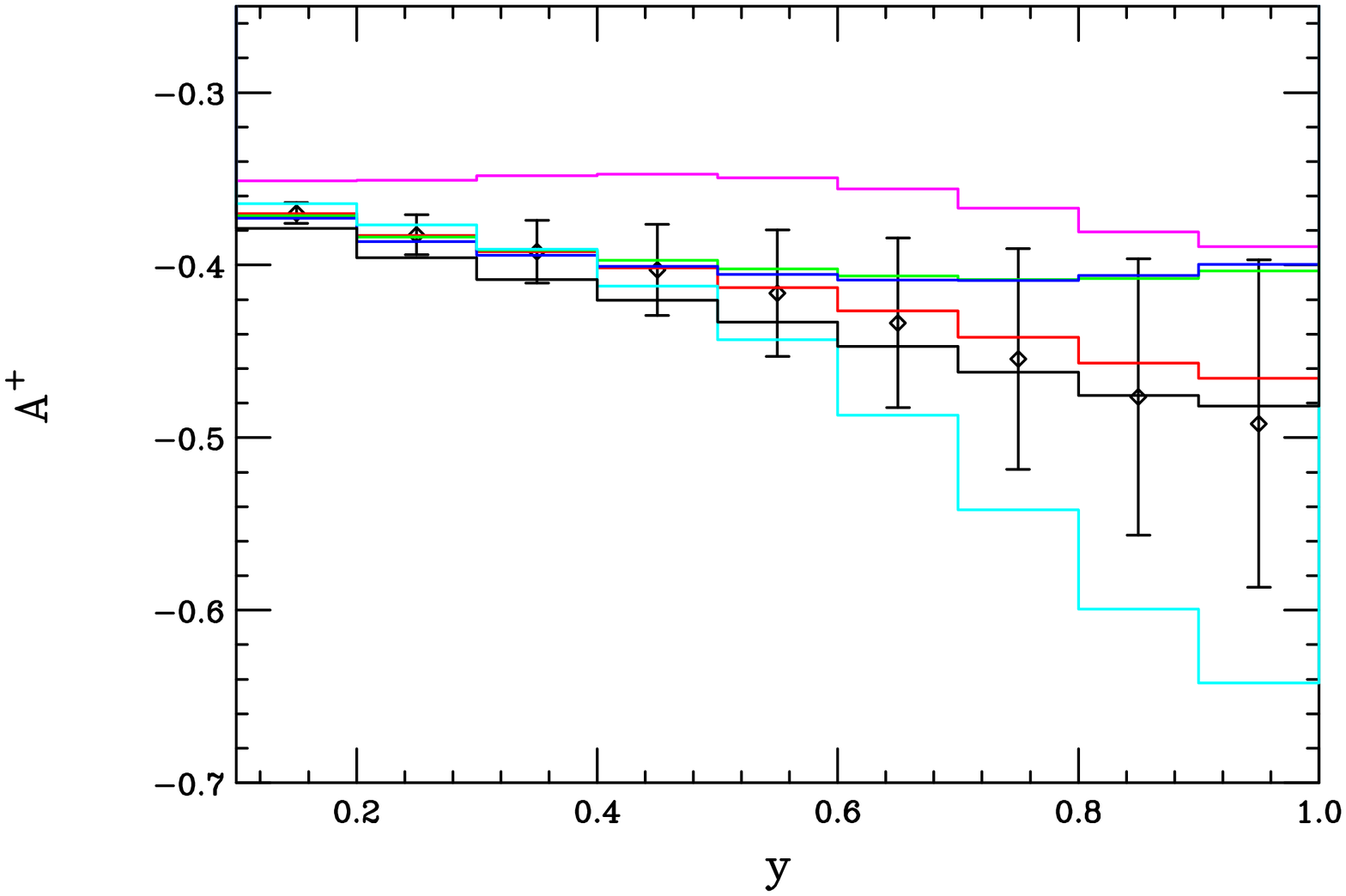}}
\vspace*{0.1cm}
\centerline{
\includegraphics[width=5.7cm,angle=90]{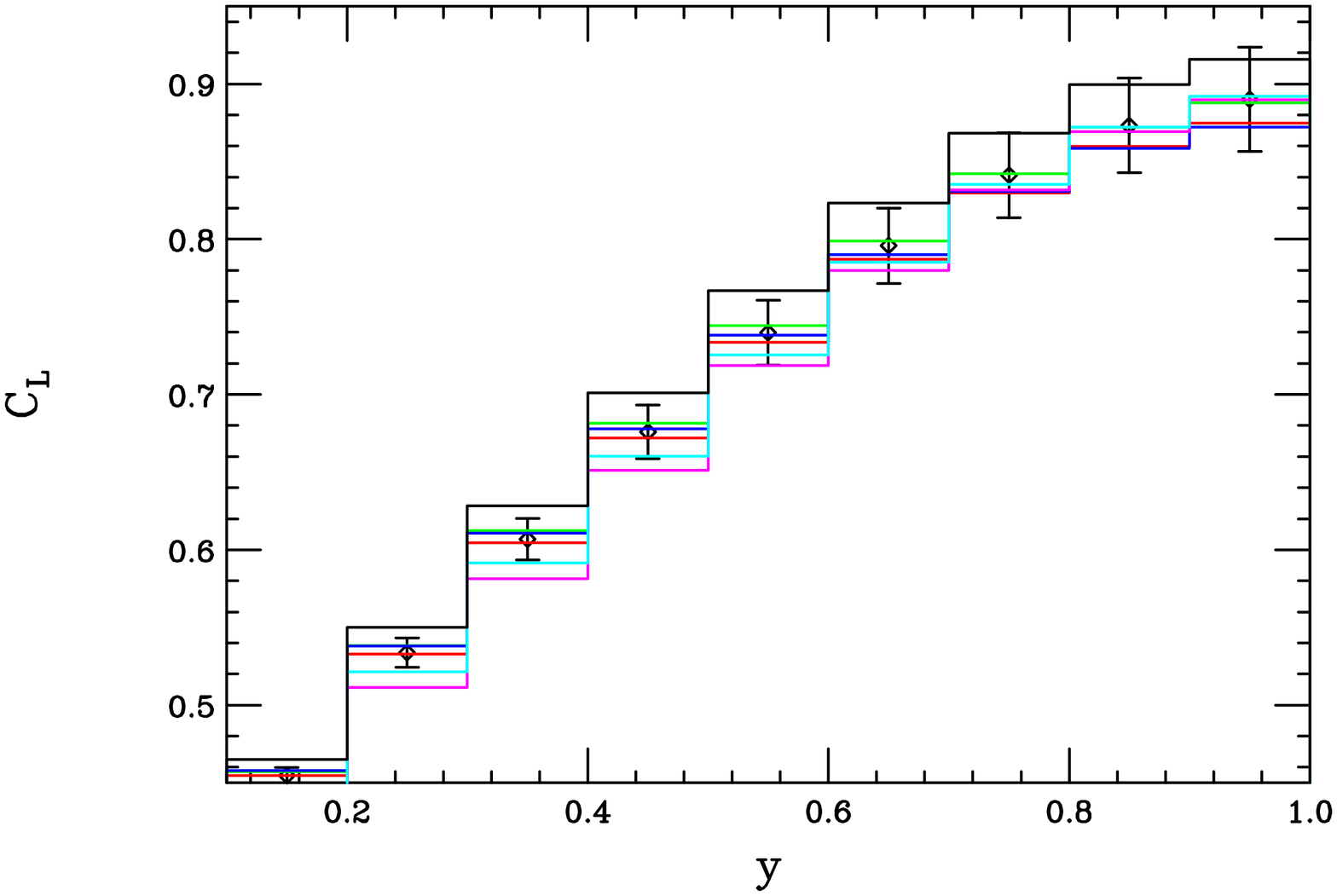}}
\caption{Same as Fig.1 but now for $M_{Z'}=1.5$ TeV.}
\label{fig4}
\end{figure}

Note further that while the $E_6$ model $\psi$ does not produce a visible signal deviating from the SM, the other models are more easily differentiated. 
In the case of the $A^-$ asymmetry, we see that the $\chi,\eta$ and SSM models lead to a larger value of this asymmetry for all $y$ while both the ALRM 
and LRM cases lead to systematically smaller values which clearly separates these two sets. This particular asymmetry shows significant model 
differentiation power; in the $A^+$ case, this power is seen to be somewhat reduced even at large $y$ 
values. In this case we see that while 
the ALRM(LRM) leads to systematically larger(smaller) values of $A^+$ (in magnitude), the other models are difficult to differentiate from the SM.
In the case of $C_L$, on the otherhand, the SSM predicts larger values than in the SM while both the ALRM and LRM cases lead to systematically 
smaller values. Similar analyses and comparisons can be made employing the other asymmetry variables. What we learn from this discussion is that the 
`best' asymmetry for differentiation with respect to the SM and the corresponding model sensitivity depends upon the nature of the couplings of the $Z'$ 
which are actually realized in nature. It is interesting to note that the charge asymmetry, $A_Q$, which does not rely on beam polarization, {\it does} 
display a strong model-dependent sensitivity which is helped by the greater statistics available since the different polarization channels can be combined.

\begin{figure}[htbp]
\centerline{
\includegraphics[width=5.7cm,angle=90]{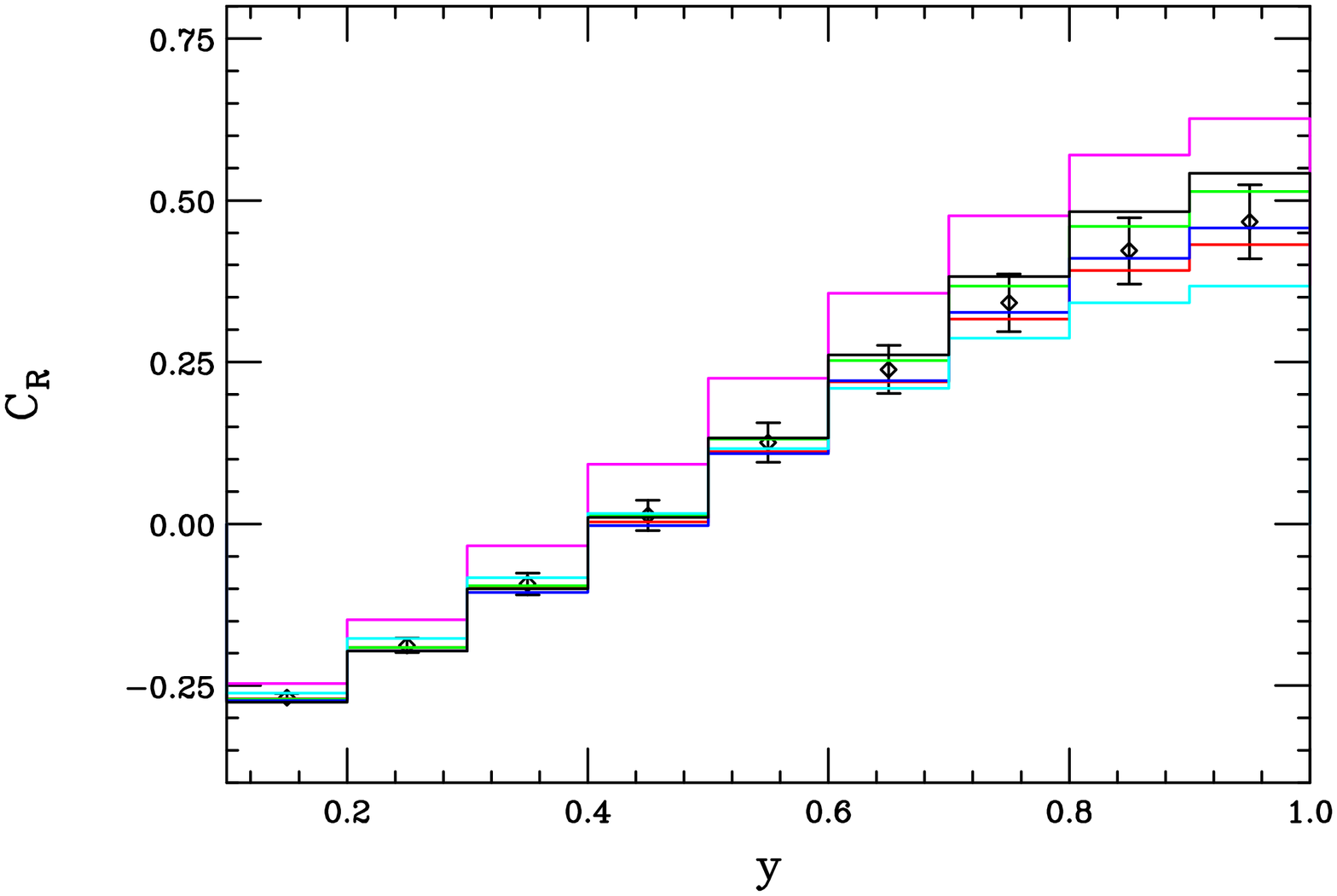}
\hspace*{0.1cm}
\includegraphics[width=5.7cm,angle=90]{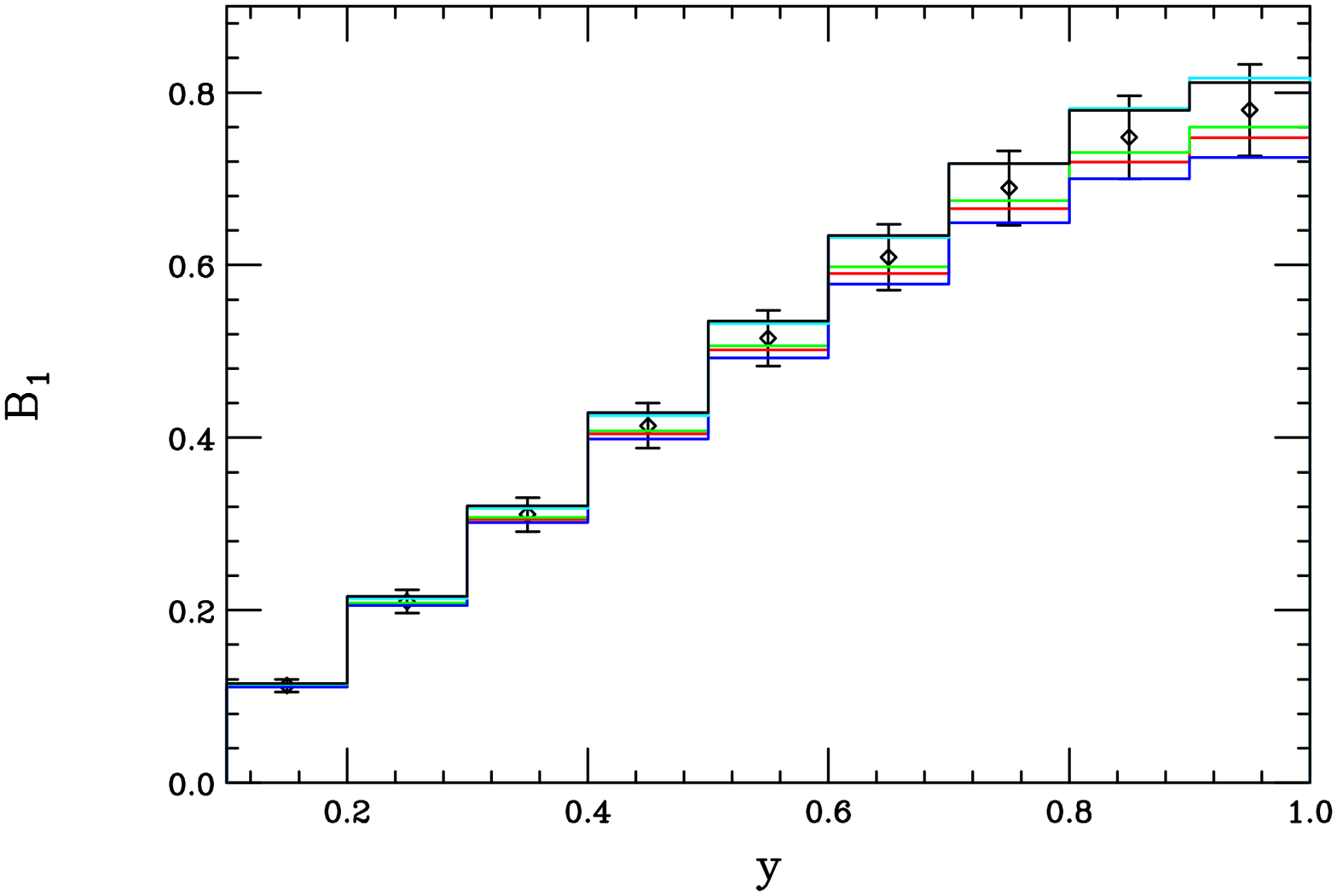}}
\vspace*{0.2cm}
\centerline{
\includegraphics[width=5.7cm,angle=90]{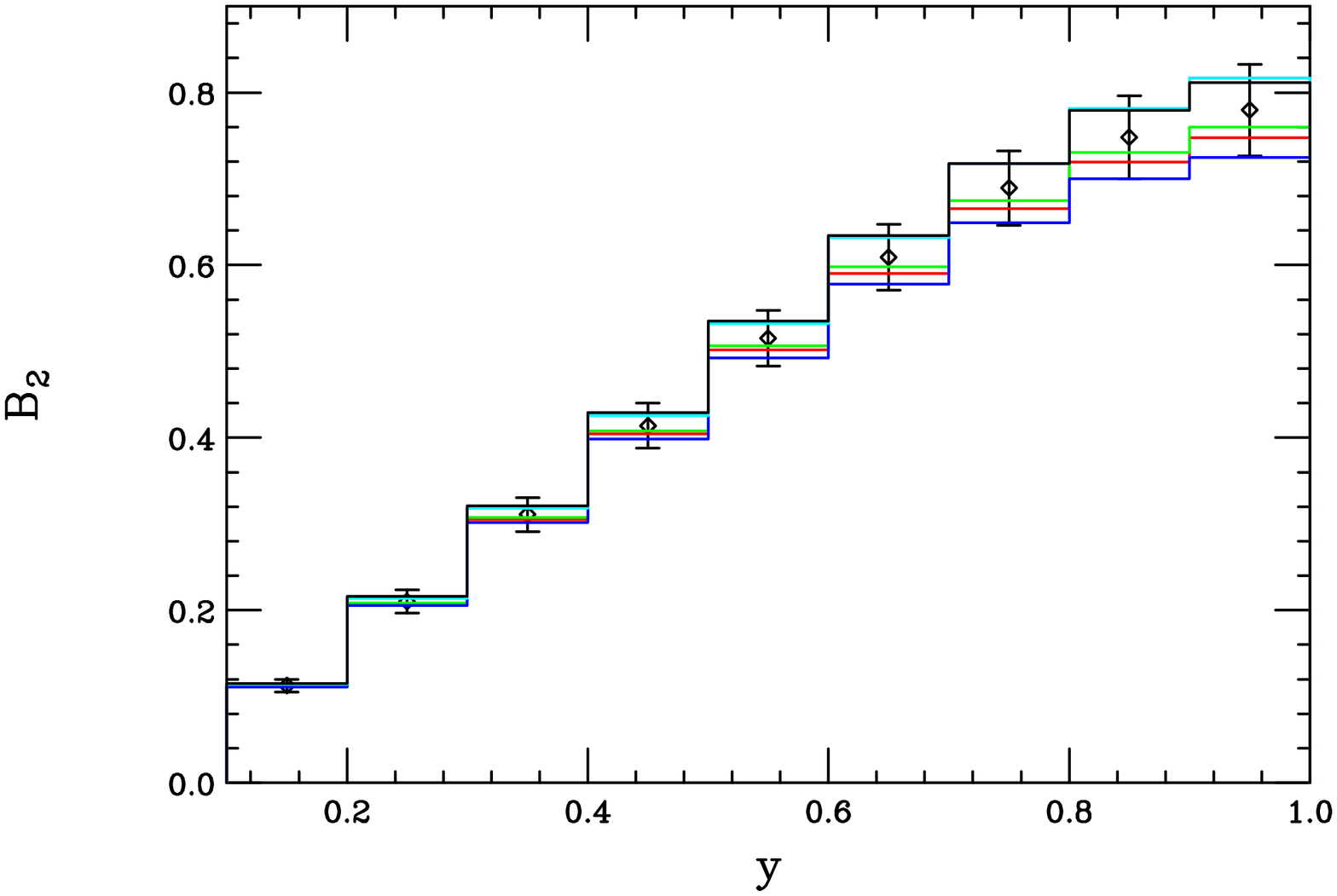}}
\caption{Same as Fig.2 but now for $M_{Z'}=1.5$ TeV.}
\label{fig5}
\end{figure}

\begin{figure}[htbp]
\centerline{
\includegraphics[width=5.7cm,angle=90]{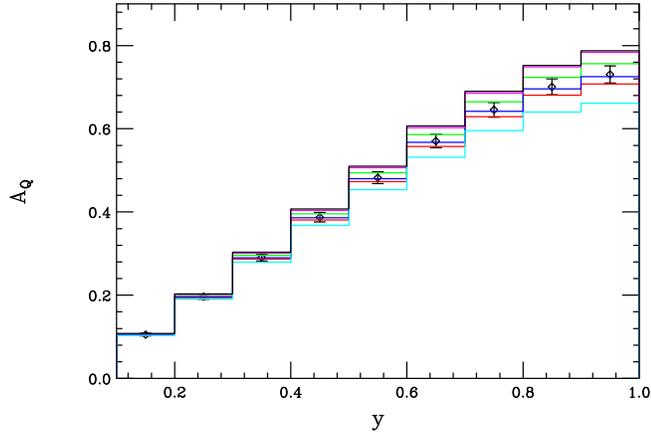}}
\caption{Same as Fig.3 but now for $M_{Z'}=1.5$ TeV.}
\label{fig6}
\end{figure}

Now let us consider a simple modification to the above scenario by raising the $Z'$ mass to 1.5 TeV from 1.2 TeV. Certainly, we 
now anticipate that the deviations from the expectations of the SM to be somewhat smaller in magnitude; this is indeed what we  
find as shown in Figs.~\ref{fig4},~\ref{fig5} and ~\ref{fig6}. What is of particular interest is the magnitude of this effect which 
is seen to be an approximate $\sim 10-40\%$ overall reduction in the size of the shift from the SM expectations depending upon the model 
and the particular choice of asymmetry examined. 
However, the overall patterns that we observed in the previous example are clearly repeated. Thus 
going to even a slightly heavier $Z'$ at fixed $\sqrt s$ and integrated luminosity can result in a substantial reduction in the 
sensitivity of these observables to the new $Z'$ couplings. It is clear from these figures that any additional increase in the $Z'$ mass would 
only further reduce the sensitivity to the $Z'$ couplings. Certainly for masses approaching 2 TeV or more measurements at these collider energies 
and integrated luminosities would not prove very useful for $Z'$ coupling extraction at least for the set of models we are examining here.
It is thus rather clear that the LHeC as currently envisioned would only be useful in this regard if the $Z'$ was rather light. 

\begin{figure}[htbp]
\centerline{
\includegraphics[width=5.7cm,angle=90]{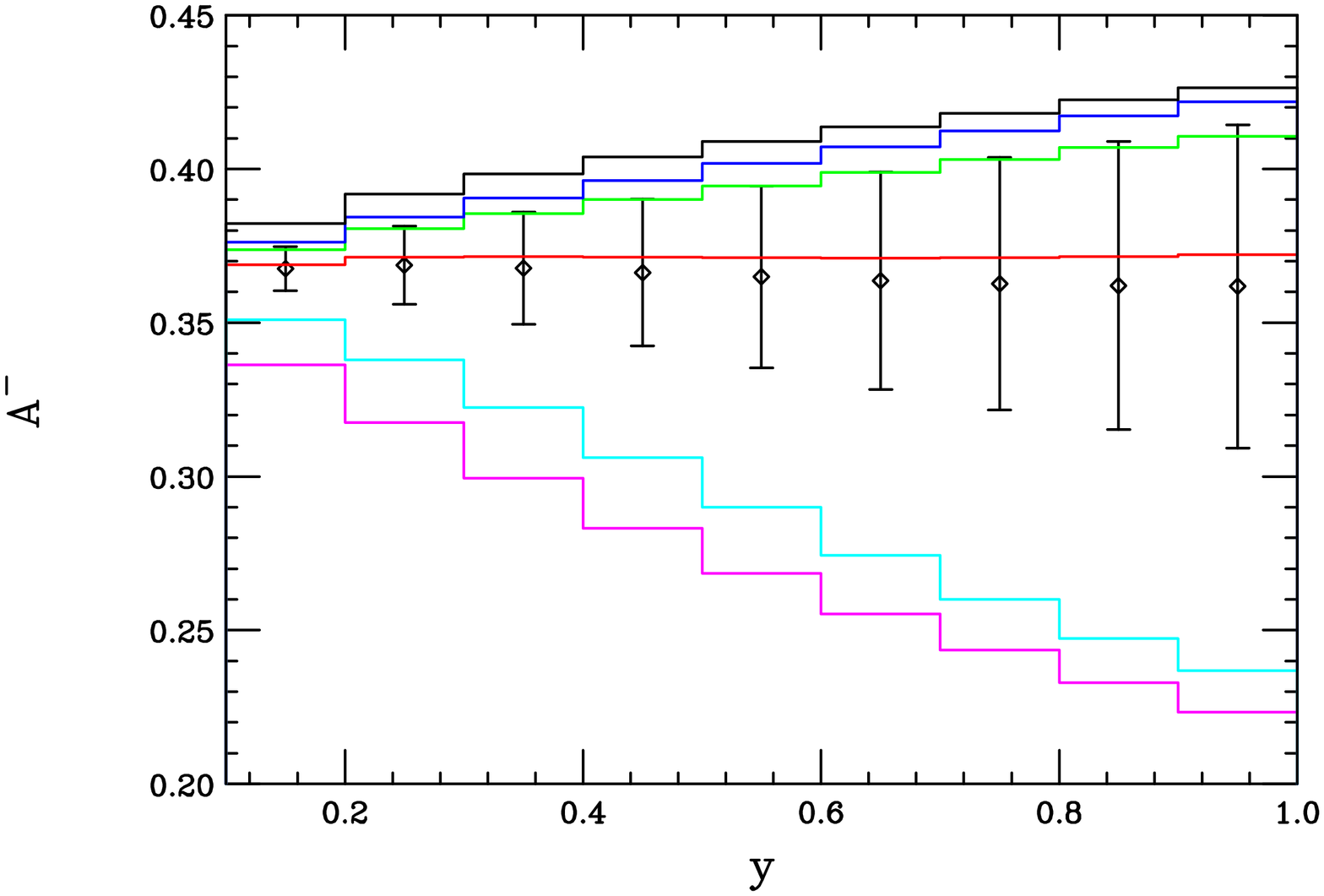}
\hspace*{0.1cm}
\includegraphics[width=5.7cm,angle=90]{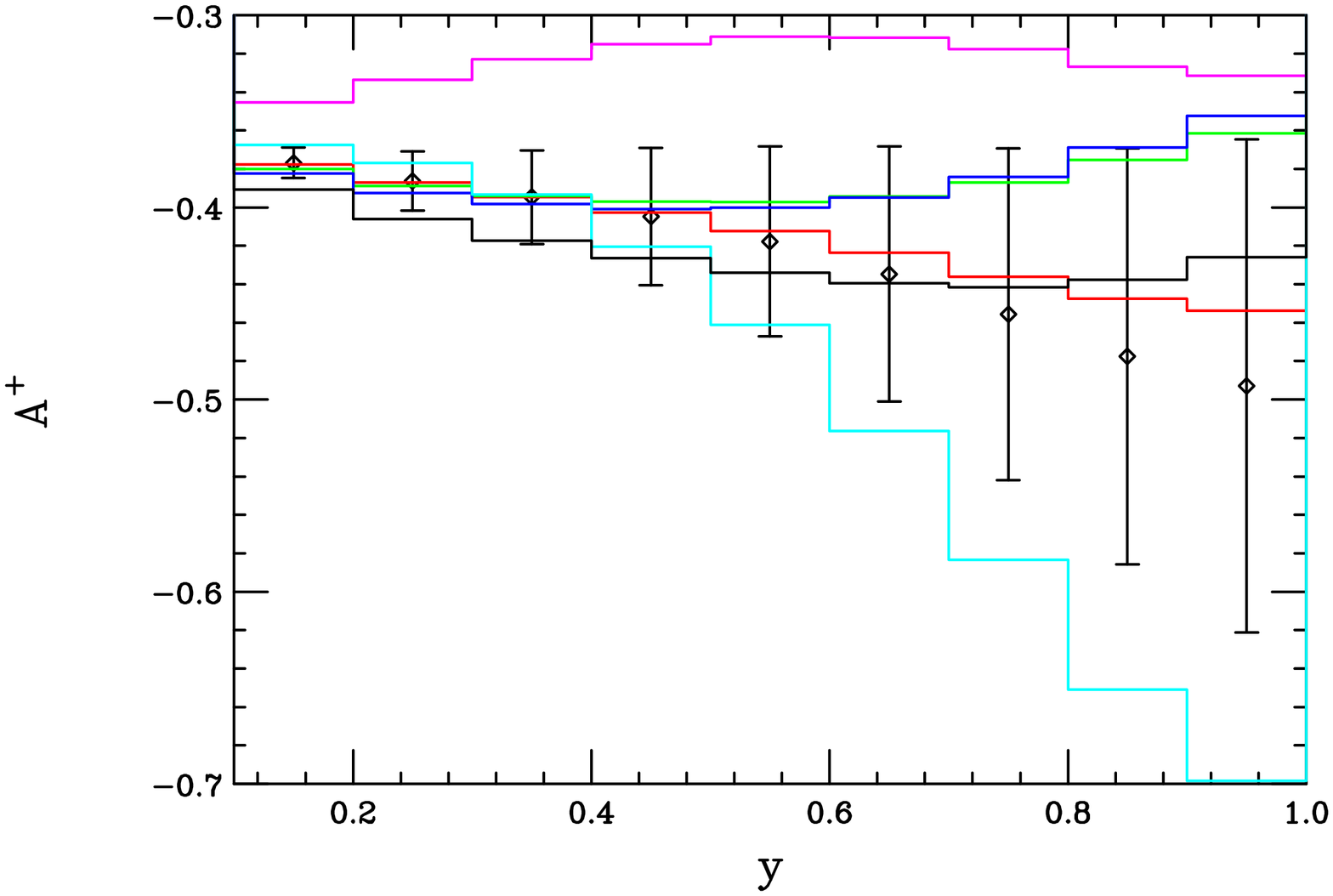}}
\vspace*{0.1cm}
\centerline{
\includegraphics[width=5.7cm,angle=90]{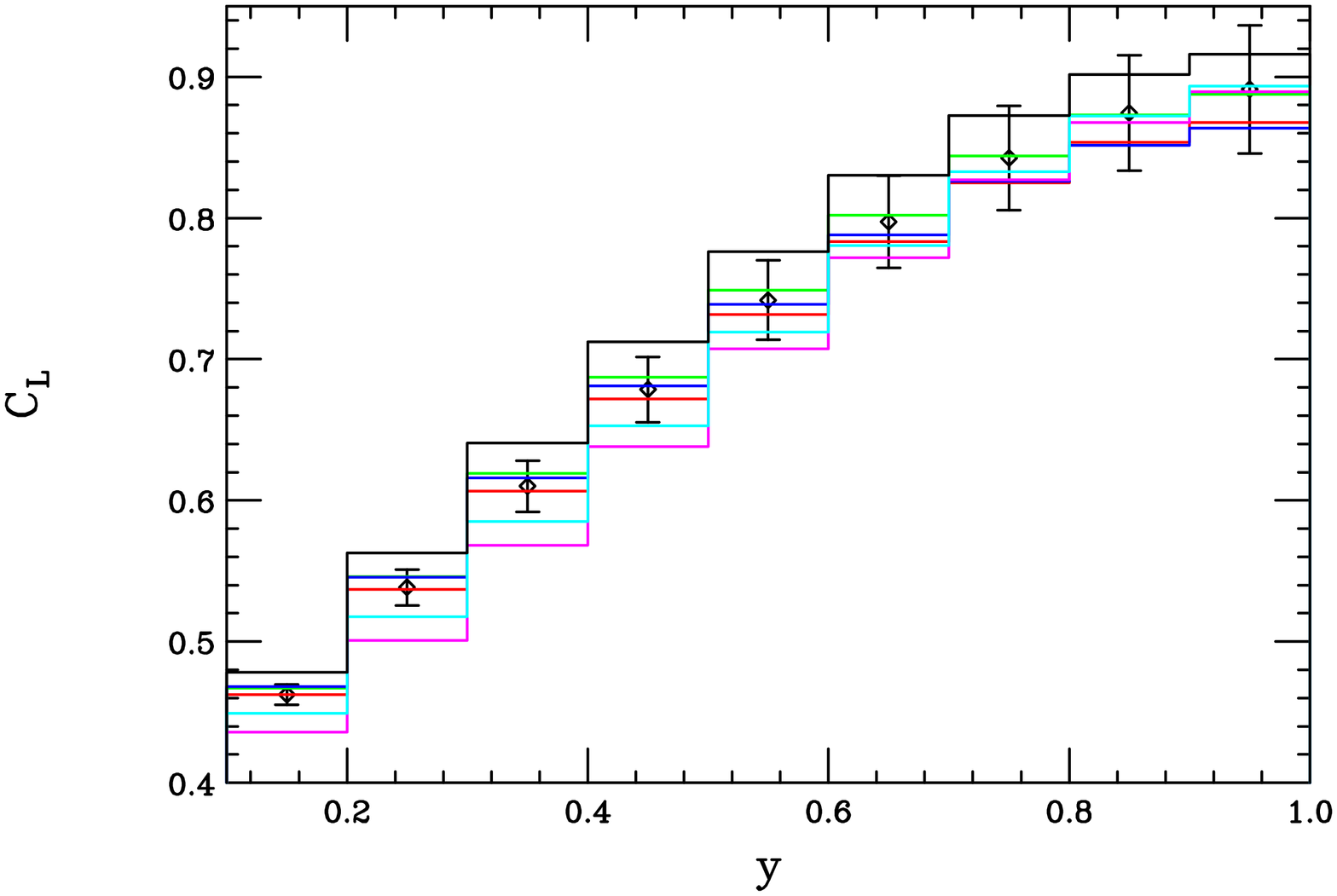}}
\caption{Same as Fig.1 but now for $M_{Z'}=1.5$ TeV with $\sqrt s=2$ TeV.}
\label{fig7}
\end{figure}

\begin{figure}[htbp]
\centerline{
\includegraphics[width=5.7cm,angle=90]{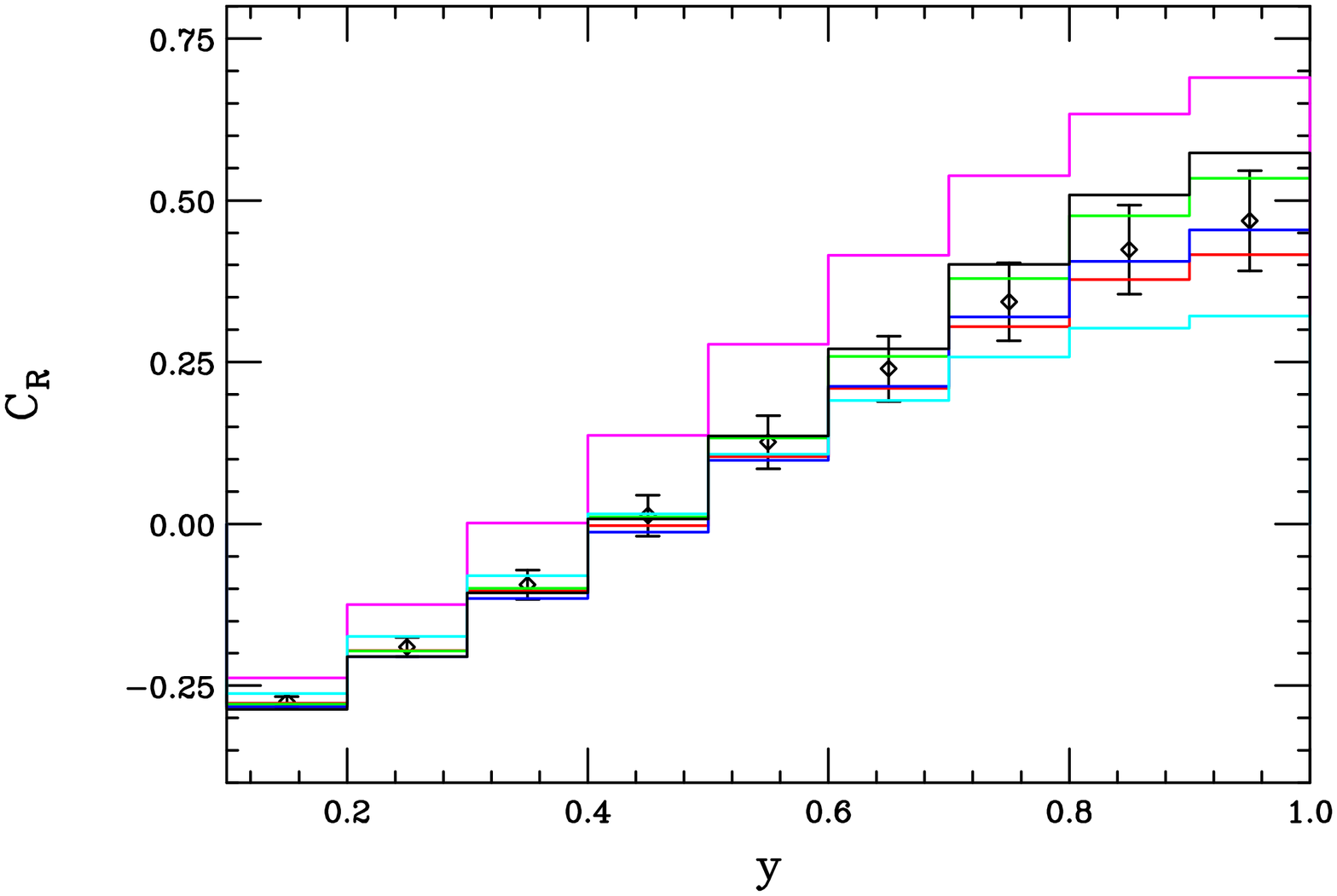}
\hspace*{0.1cm}
\includegraphics[width=5.7cm,angle=90]{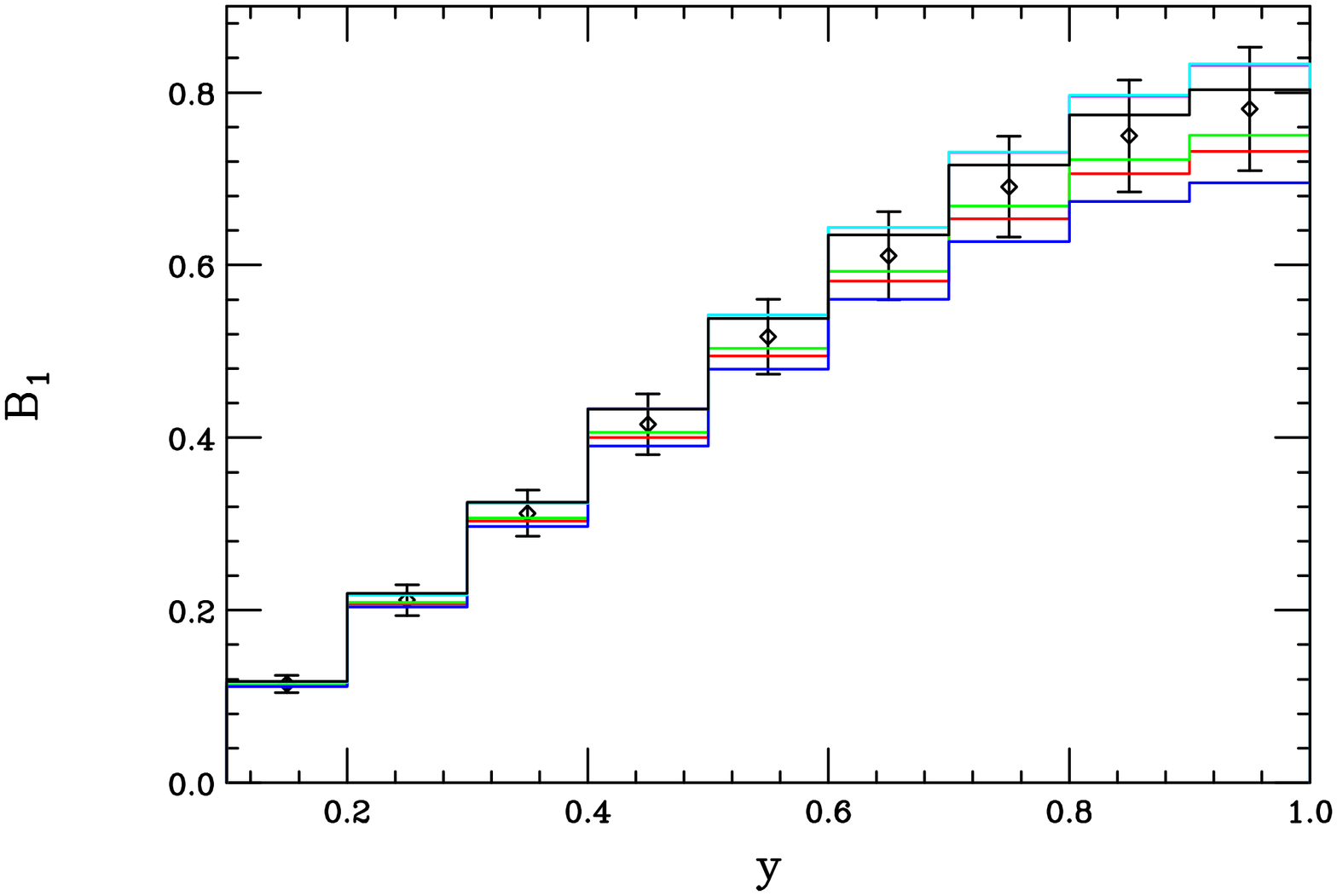}}
\vspace*{0.2cm}
\centerline{
\includegraphics[width=5.7cm,angle=90]{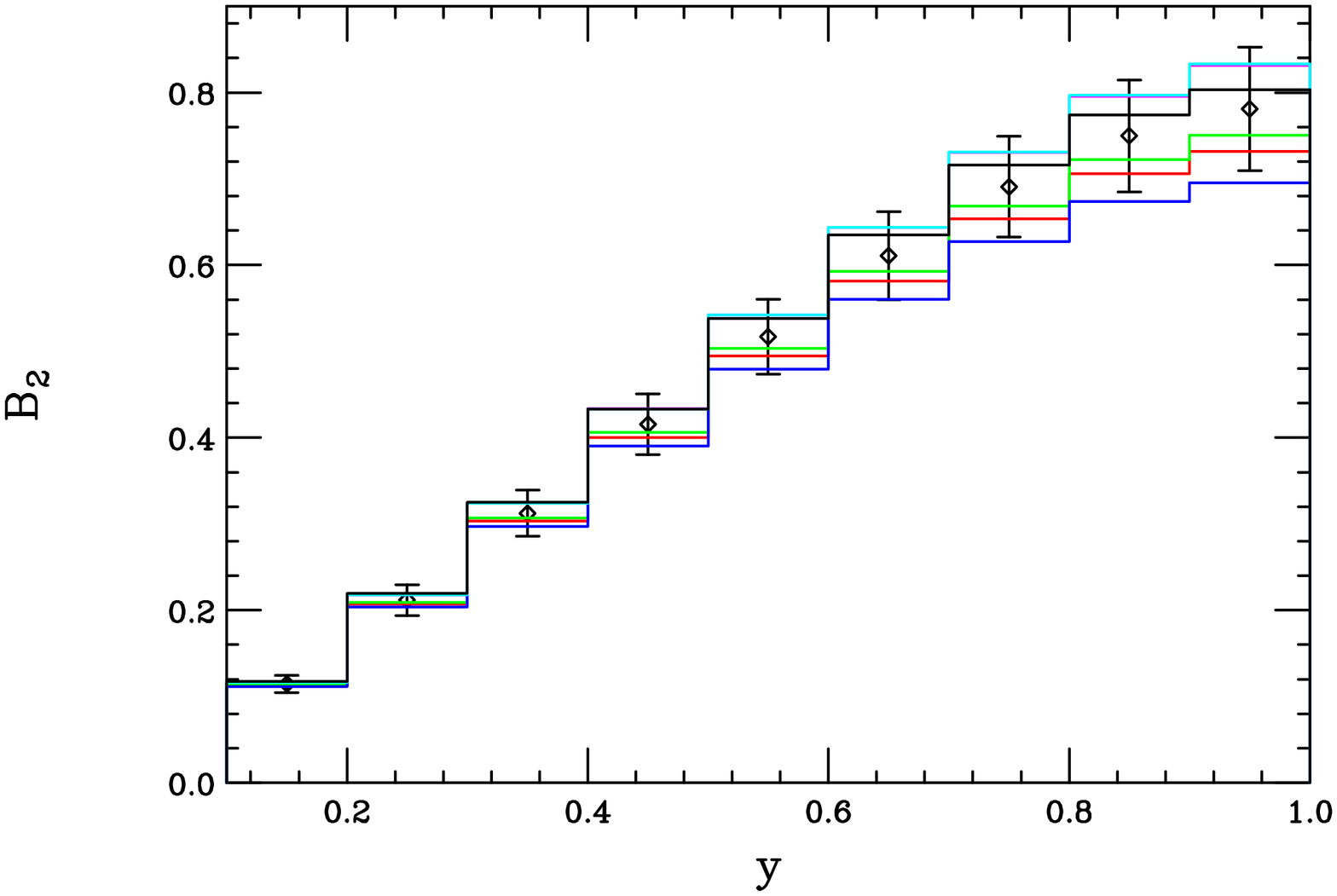}}
\caption{Same as Fig.2 but now for $M_{Z'}=1.5$ TeV with $\sqrt s=2$ TeV.}
\label{fig8}
\end{figure}

\begin{figure}[htbp]
\centerline{
\includegraphics[width=5.7cm,angle=90]{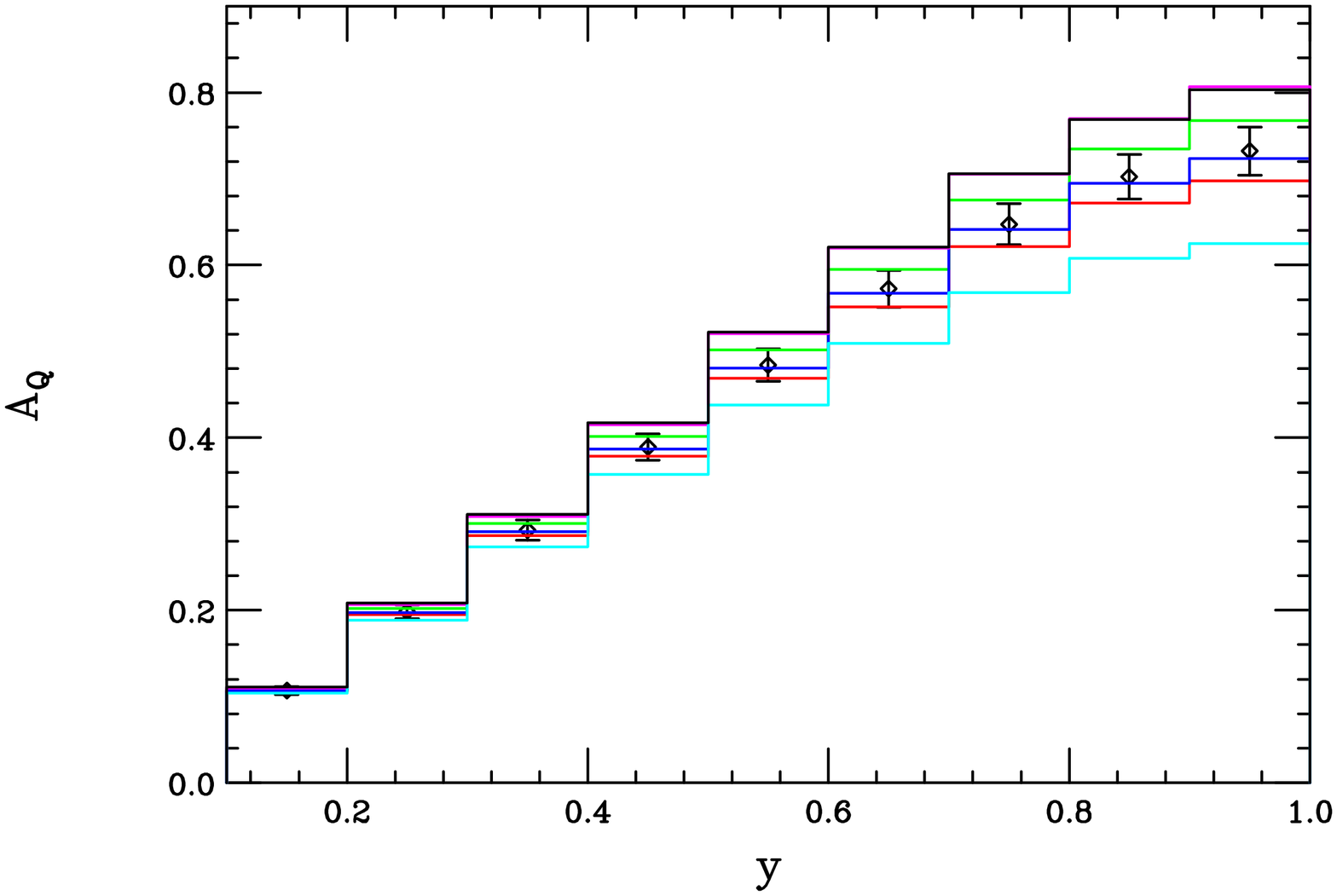}}
\caption{Same as Fig.3 but now for $M_{Z'}=1.5$ TeV with $\sqrt s=2$ TeV.}
\label{fig9}
\end{figure}

Can we recover any of this lost sensitivity by going to even larger values of $\sqrt s$? To address this issue let us now consider the scenario  
where $\sqrt s=2$ TeV and $M_{Z'}=1.5$ TeV. There are two competing factors entering in this case 
in comparison to those previously considered. 
First, going to a larger value of $\sqrt s$ for a fixed set of cuts on the variables $x,y$ implies a larger minimum value of $Q^2$ which thus 
increases the sensitivity to $Z'$ exchange. However, there is a second effect due to the $1/s$ scaling behavior of the differential cross section 
which leads to a lower event rate for a fixed value of the integrated luminosity. Thus while Figs.~\ref{fig7},~\ref{fig8} and ~\ref{fig9} do show 
an increased sensitivity to $Z'$ exchange due to the higher $Q^2$ values being probed, the error bars on the SM expectations are also increased in 
size due to the lower event rates. It would appear that the power of the $\sqrt s=2$ TeV and $M_{Z'}=1.5$ TeV case to probe $Z'$ couplings thus lies 
in the intermediate range between the $\sqrt s=1.5 $ TeV with $M_{Z'}=1.2$ TeV and the $\sqrt s=M_{Z'}=1.5$ TeV cases discussed above.


\section{Discussion and Conclusions}

In this paper we have begun to explore the capability of a hybrid $ep$ collider upgrade at the LHC, the LHeC, to probe the couplings of 
a hypothetical light $Z'$ in the mass range below $\lsim 1.5$ TeV in order to complement those that would be made in the LHC Drell-Yan channel. 
Such a collider would be expected 
to operate at a center of mass energy of $\sqrt s=1.5-2$ TeV with $e^\pm$ beam polarization and may collect an integrated luminosity 
of order 100 $fb^{-1}$. The LHeC may provide the only means to obtain additional information on the $Z'$ couplings until a linear 
collider is constructed. Although the details of a possible LHeC project are not yet firm, the analysis above indicates that asymmetry measurements 
at these energies are quite sensitive to the specific values of the $Z'$ couplings in a number of common representative models.   
This information, when combined with that obtainable directly from more typical LHC data, may allow for a complete determination of the 
couplings of such a new gauge boson in a model-independent way, something which is impossible to do with data from the LHC alone for a $Z'$ in 
this mass range. However, if the $Z'$ were to be more massive than about $\sim 1.5-2$ TeV it is unlikely that such an $ep$ collider would be of much 
use in this regard. Fortunately, we  will know relatively early on in the running of the LHC whether or not a $Z'$ in this mass reach does indeed 
exist. To more fully address the issues raised here a significantly more detailed study will become necessary (if required by the data) once the 
parameters of the LHeC become definitive. Hopefully a $Z'$ will be observed in early LHC running.

%
%%%%%%%%%%%%%%%%%%--- References
%%%%%%%%%%%%%%%%%%%%%%%%%%%%%%%%%%%%%%%%%%%%%%%%%%%%%%%
\def\MPL #1 #2 #3 {Mod. Phys. Lett. {\bf#1},\ #2 (#3)}
\def\NPB #1 #2 #3 {Nucl. Phys. {\bf#1},\ #2 (#3)}
\def\PLB #1 #2 #3 {Phys. Lett. {\bf#1},\ #2 (#3)}
\def\PR #1 #2 #3 {Phys. Rep. {\bf#1},\ #2 (#3)}
\def\PRD #1 #2 #3 {Phys. Rev. {\bf#1},\ #2 (#3)}
\def\PRL #1 #2 #3 {Phys. Rev. Lett. {\bf#1},\ #2 (#3)}
\def\RMP #1 #2 #3 {Rev. Mod. Phys. {\bf#1},\ #2 (#3)}
\def\NIM #1 #2 #3 {Nuc. Inst. Meth. {\bf#1},\ #2 (#3)}
\def\ZPC #1 #2 #3 {Z. Phys. {\bf#1},\ #2 (#3)}
\def\EJPC #1 #2 #3 {E. Phys. J. {\bf#1},\ #2 (#3)}
\def\IJMP #1 #2 #3 {Int. J. Mod. Phys. {\bf#1},\ #2 (#3)}
\def\JHEP #1 #2 #3 {J. High En. Phys. {\bf#1},\ #2 (#3)}

\end{document}